\documentstyle[hx37_latex]{article}

\def\be{\begin{equation}}
\def\ee{\end{equation}}
\def\bea{\begin{eqnarray}}
\def\eea{\end{eqnarray}}

\def\plotfiddle#1#2#3#4#5#6#7{\centering \leavevmode
    \vbox to#2{\rule{0pt}{#2}}
    \includegraphics{#1}}

\begin{document}

{ \hyphenpenalty=5000 \noindent
To be published in {\it HST and the High Redshift Universe (37$^{\rm th}$
Herstmonceux Conference)}, eds.  N.R. Tanvir, A. Aragon-Salamanca,
J.V. Wall, 1996. } 

\vspace{1.5cm}

\title{THE ULTRAVIOLET MORPHOLOGY OF GALAXIES}

\author{ ROBERT W. O'CONNELL, PAMELA MARCUM}

\address{University of Virginia,\\ Charlottesville, VA 22903-0818, USA}

\def\lesssim{\mathrel{\hbox{\rlap{\hbox{\lower4pt\hbox{$\sim$}}}\hbox{$<$}}}}
\def\gtrsim{\mathrel{\hbox{\rlap{\hbox{\lower4pt\hbox{$\sim$}}}\hbox{$>$}}}}
\def\arcsec{\hbox{$^{\prime\prime}$}}
\def\arcmin{\hbox{$^{\prime}$}}

\maketitle\abstracts{
Optical band images of distant (z$\,\gtrsim\, 0.5$)
galaxies, such as those of the Hubble Deep Field, record light from
the rest-frame vacuum ultraviolet ($\lambda < 3000$ \AA).  Because the
appearance of a galaxy is a very strong function of wavelength, and
especially so in the UV, evolutionary studies of distant galaxies can
be seriously influenced by a ``morphological k-correction''
effect.  We use images obtained by the Ultraviolet Imaging Telescope
during the {\it Astro} missions to explore the extent of
this effect  and intercompare far-UV with optical morphologies
for various types of galaxies.  }


The morphology of a distant galaxy is often our first index to its
evolutionary state.  Unfortunately, there is a host of technical
complications which afflict the apparent morphologies of distant
galaxies.  These can conspire to make the familiar look unfamiliar and
vice-versa.  

Two of these difficulties are well known:  {\it (i)} reduced spatial
resolution; and {\it (ii)} the rapid fading of surface brightness with
redshift, $SB \sim (1+z)^{-n}$, where n = 3--5, depending on the
method used to characterize surface brightness.  These two effects are
relatively simple to model, and their consequences for ground-based
and HST imaging have recently received much
attention~\cite{dri95}~\cite{lil95}~\cite{ab96a}~\cite{dr96}~\cite{gia96}.  
Here, we wish to focus on a third difficulty, which can be called the
``morphological k correction''---i.e.~the fact that the appearance of
galaxies is a strong function of wavelength.  The photons recorded on
exposures of distant galaxies in optical bands, e.g.~in the Hubble
Deep Field, often originate in the vacuum ultraviolet in the
restframe.  It has been known for many years that the appearance of
galaxies in the UV can be very different than at the optical
wavelengths for which the classical morphological typing systems were
established~\cite{boh83}~\cite{kingell}.  In the UV, hot stars are
emphasized and cool ones are suppressed.  This provides an important
opportunity to distinguish stellar populations, but it also means that
morphological analyses must take into account a strong UV/optical
k-correction.  

It has been hard to quantify this effect because few vacuum-UV images
of nearby galaxies were available before 1990.  Since then, several
hundred objects have been observed in the UV by HST, the FOCA balloon
experiment~\cite{blech}, and the Ultraviolet Imaging
Telescope~\cite{stech} (UIT) on the two {\it Astro} missions.  Maos et
al.~\cite{maoz96} have recently published an atlas of 110 nearby galaxy
nuclei observed at 2300 \AA\ in the 22\arcsec\ field of the HST Faint
Object Camera.  We~\cite{atlas} are preparing a UV/optical atlas which
compares the UV and optical morphologies of 27 nearby galaxies using
UIT data mainly from the {\it Astro-1} mission.  A later paper will
add 45 galaxies observed during {\it Astro-2}.  UIT's field of
view is 40\arcmin\ in diameter and produced typical resolution of
3\arcsec\ FWHM; the two principal (solar-blind) imaging bands were
centered at 1500 \AA\ (``far-UV'') and 2500 \AA\ (``mid-UV'').  UIT's
characteristics are well suited to the study of morphologies of nearby
galaxies.  Some of this data has already been used to simulate the
combined effects of the k-correction, reduced resolution, and reduced
surface brightness on observations of distant
galaxies~\cite{gia96}~\cite{boh91}.

Here we show a few of the comparisons from the UIT Atlas and briefly
mention the effects they illustrate.  Data for a number of spirals
shows that dust in normal disks does not strongly affect UV
morphologies (Fig.~1).  However, spirals do shift to later Hubble
types in the UV (Fig.~2).  The k-correction combined with reduced
resolution and bias toward high SB regions can turn a single normal
galaxy into an irregular object with ``interacting'' companions
(Fig.~3).  Disturbed dust layers can have drastic effects on UV light
(e.g.~M82, not shown); for instance, an interacting pair of galaxies
can turn into an isolated galaxy (Fig.~4).  Barred galaxies often
become unbarred (e.g.~M83~\cite{boh83}).  The general trend found is
from normal to abnormal.  To put this another way, the incidence of
abnormal galaxies in the restframe UV appears to be larger than it
really is.

\begin{figure} 
\plotfiddle{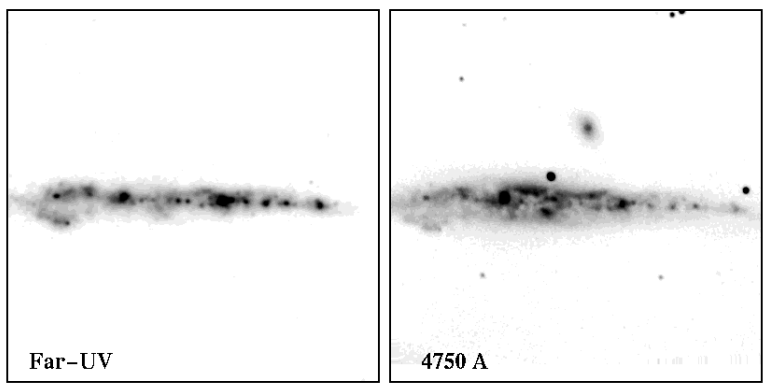}{2.0truein}{0}{90}{90}{-148}{-45}
\caption{Dust does not dominate the UV appearance of normal disk galaxies, even
in edge-on systems like this one.}
\end{figure}

\begin{figure}
\plotfiddle{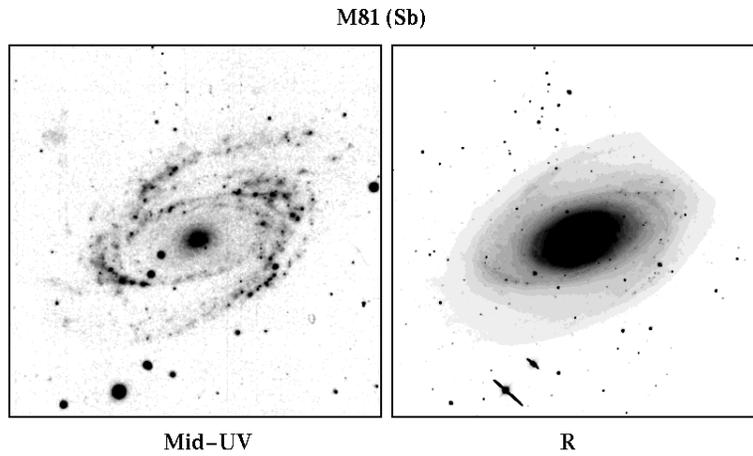}{2.0truein}{0}{90}{90}{-148}{-40}
\caption{Hubble types for spirals become later in the UV. In the far-UV
(not shown), M81 becomes a nearly ``empty'' ring.}
\end{figure}

\begin{figure}
\plotfiddle{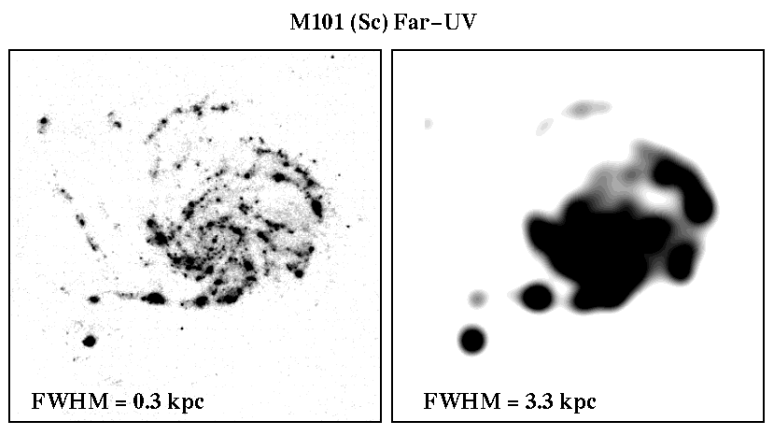}{2.0truein}{0}{90}{90}{-148}{-45}
\caption{Effects of k-correction and reduced resolution, corresponding
to observations at $z \sim 1$ with 0.4\arcsec\ FWHM.  Bright associations
have become ``companions''.}
\end{figure}

\begin{figure}
\plotfiddle{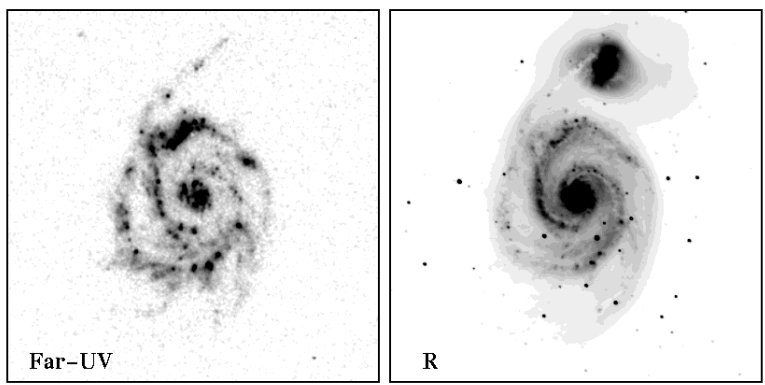}{2.0truein}{0}{90}{90}{-148}{-45}
\caption{Case of a disappearing perturber.}
\end{figure}


\end{document}